\documentclass{emulateapj}
\usepackage{amssymb,bm,graphicx,color}
\newcommand{\asca}{{\itshape ASCA}}
\newcommand{\chandra}{{\itshape Chandra}}
\newcommand{\xmm}{{\itshape XMM-Newton}}
\newcommand{\suzaku}{{\itshape Suzaku}}
\newcommand{\rate}{counts s$^{-1}$}
\newcommand{\lum}{ergs s$^{-1}$}
\newcommand{\flux}{ergs cm$^{-2}$ s$^{-1}$}
\newcommand{\SB}{ergs cm$^{-2}$ s$^{-1}$ sr$^{-1}$}
\newcommand{\edens}{ergs cm$^{-3}$}
\newcommand{\colden}{cm$^{-2}$}
\newcommand{\ue}{$u_{\rm e}$}
\newcommand{\um}{$u_{\rm m}$}
\newcommand{\src}{3C~35}
\newcommand{\rxs}{1RXS J$011243.5$+$492930$}

\slugcomment{}
\shorttitle{Energetics in the lobes of 3C 35}
\shortauthors{N. Isobe et al.}

\begin{document}
%% Title %%%
\title{\suzaku~diagnostics of the energetics
in the lobes of the giant radio galaxy \src}

%% authors %%
\author{ 
Naoki      Isobe       \altaffilmark{1}, 
Hiromi     Seta        \altaffilmark{2},
Poshak     Gandhi      \altaffilmark{3}, \&
Makoto S.  Tashiro     \altaffilmark{2} 
}

\altaffiltext{1}{Department of Astronomy, Kyoto University, 
        Kitashirakawa-Oiwake-cho, Sakyo-ku, Kyoto 606-8502, Japan}
\email{n-isobe@kusastro.kyoto-u.ac.jp}
\altaffiltext{2}{Department of Physics, Saitama University,
        255 Shimo-Okubo, Sakura-ku, Saitama, 338-8570, Japan}
\altaffiltext{3}{Department of High Energy Astrophysics, ISAS/JAXA,  
3-1-1 Yoshinodai, Chuo-ku, Sagamihara, Kanagawa 229-8510, Japan}

\begin{abstract}
The \suzaku~observation of a giant radio galaxy \src~revealed 
faint extended X-ray emission,
associated with its radio lobes and/or host galaxy.
After careful subtraction of the X-ray and non-X-ray background 
and contaminating X-ray sources, 
the X-ray spectrum of the faint emission was reproduced 
by a sum of the power-law (PL) and soft thermal components. 
The soft component was attributed 
to the thermal plasma emission from the host galaxy. 
The photon index of the PL component, 
$\Gamma = 1.35_{-0.86}^{+0.56}$$_{-0.10}^{+0.11}$ 
where the first and second errors represent 
the statistical and systematic ones, 
was found to agree with the synchrotron radio index 
from the lobes, $\Gamma_{\rm R} = 1.7$.
Thus, the PL component was attributed to the inverse Compton (IC) X-rays 
from the synchrotron electrons in the lobes.
The X-ray flux density at 1 keV was derived as $13.6\pm 5.4_{-3.6}^{+4.0}$ nJy 
with the photon index fixed at the radio value.
The X-ray surface brightness from these lobes ($\sim 0.2$ nJy arcmin$^{-2}$) 
is lowest among the lobes studied through the IC X-ray emission. 
In combination with the synchrotron radio flux density, 
$7.5 \pm 0.2$ Jy at 327.4 MHz, 
the electron energy density spatially averaged over the lobes
was evaluated to be the lowest among those radio galaxies,
as $u_{\rm e} = (5.8 \pm 2.3 _{-1.7}^{+1.9}) \times 10^{-14}$ ergs cm$^{-3}$ 
over the electron Lorentz factor of $10^{3}$ -- $10^{5}$. 
The magnetic energy density was calculated as 
$u_{\rm m}=(3.1_{-1.0}^{+2.5}$$_{-0.9}^{+1.4}) \times 10^{-14}$ ergs cm$^{-3}$,
corresponding to the magnetic field strength of 
$0.88_{-0.16}^{+0.31}$$_{-0.14}^{+0.19}$ $\mu$G.  
These results suggest that the energetics in the \src~lobes
are nearly consistent with 
equipartition between the electrons and magnetic fields.
\end{abstract}

%%%%%%%%%%%%%
% Section 1 %
%%%%%%%%%%%%%
\section{Introduction} %===================================================
\label{sec:intro}
The lobes of radio galaxies and quasars
accumulate the kinetic energy of their jets,
after it is converted to those of the particles and magnetic fields
through the terminal hot spots. 
Accordingly, the lobes are observed 
as sources of strong synchrotron radio emission,
produced by interplay between the relativistic electrons and magnetic fields. 
These electrons inevitably emit X-ray and $\gamma$-ray photons,  
by inverse Compton (IC) scattering off 
the cosmic microwave background (CMB) radiation \citep{CMB_IC}.
A comparison of the IC X-ray and synchrotron radio intensities
successfully disentangles 
the energy densities of the electrons and magnetic fields, 
\ue~and \um~, in the lobes, 
which provide a clue to the past activity of the jets and nucleus,
and their evolution.   

Since the discovery of the IC X-ray emission from the lobes of Fornax A 
with \asca~\citep[e.g.,][]{ForA_ASCA} and {\itshape ROSAT} \citep{ForA_ROSAT},
tremendous progress was achieved 
by this method with \chandra~\citep[e.g.,][]{3C452,lobes_Croston}
and \xmm~\citep[e.g.,][]{3C98,ForA,PicA_XMM}, 
on the energetics in the lobes of radio galaxies. 
A number of studies \citep[e.g.,][]{3C452,lobes_Croston} suggested 
an electron dominance of $u_{\rm e} / u_{\rm m} = 1$ -- $100$ in the lobes. 
Correspondingly, the magnetic field was found to be weaker 
than that expected from the minimum energy condition by several factors. 

Among radio galaxies, 
those with lobes evolved to a linear size larger than $ D \sim  1$ Mpc 
are usually called as {\itshape giant radio galaxies}. 
Based on the spectral aging technique, 
they are suggested to be relatively old sources 
\citep[a typical spectral age of $\sim 100$ Myr;][]{giants_age}.
Therefore, the giant radio galaxies are utilized to probe a late phase 
in the evolution of jets and radio sources.
However, the current IC X-ray studies on the lobe energetics 
are concentrated on smaller sources 
with a total dimension of $D = 50$ -- $500$ kpc, 
and only a few IC X-ray results are reported 
for giant radio galaxies \citep[e.g.,][]{3C457_XMM}.

For the X-ray study on the extended sources, 
including giant radio galaxies 
(e.g., $\sim 1$ Mpc corresponding to $\sim 10'$ at the redshift of $z = 0.1$),
the X-ray Imaging Spectrometer \citep[XIS;][]{XISpaper} onboard
\suzaku~\citep{Suzaku} is regarded as the best observatory,
thanks to its low and stable background level \citep{xisnxbgen},
in combination with its large effective area up to 10 keV. 
By taking advantage of these capabilities, 
\citet{3C326} have made a successful detection of the IC X-rays 
with \suzaku~from the west lobe of the giant radio galaxy 3C 326
with a size of $D\sim 2$ Mpc.
As a result, 
consistent with lobes of the smaller radio galaxies,
the electron dominance of $u_{\rm e}/u_{\rm m} \sim 20$ \citep{3C326} 
was revealed in this lobe. 
This result motivated a systematic X-ray study 
with \suzaku~on the lobe energetics
to be extended to the lobes of the giant radio galaxies.

%==========% 
% Figure 1 %
%==========%
\begin{figure*}[hp]
\begin{center}
\includegraphics[width=8.0cm]{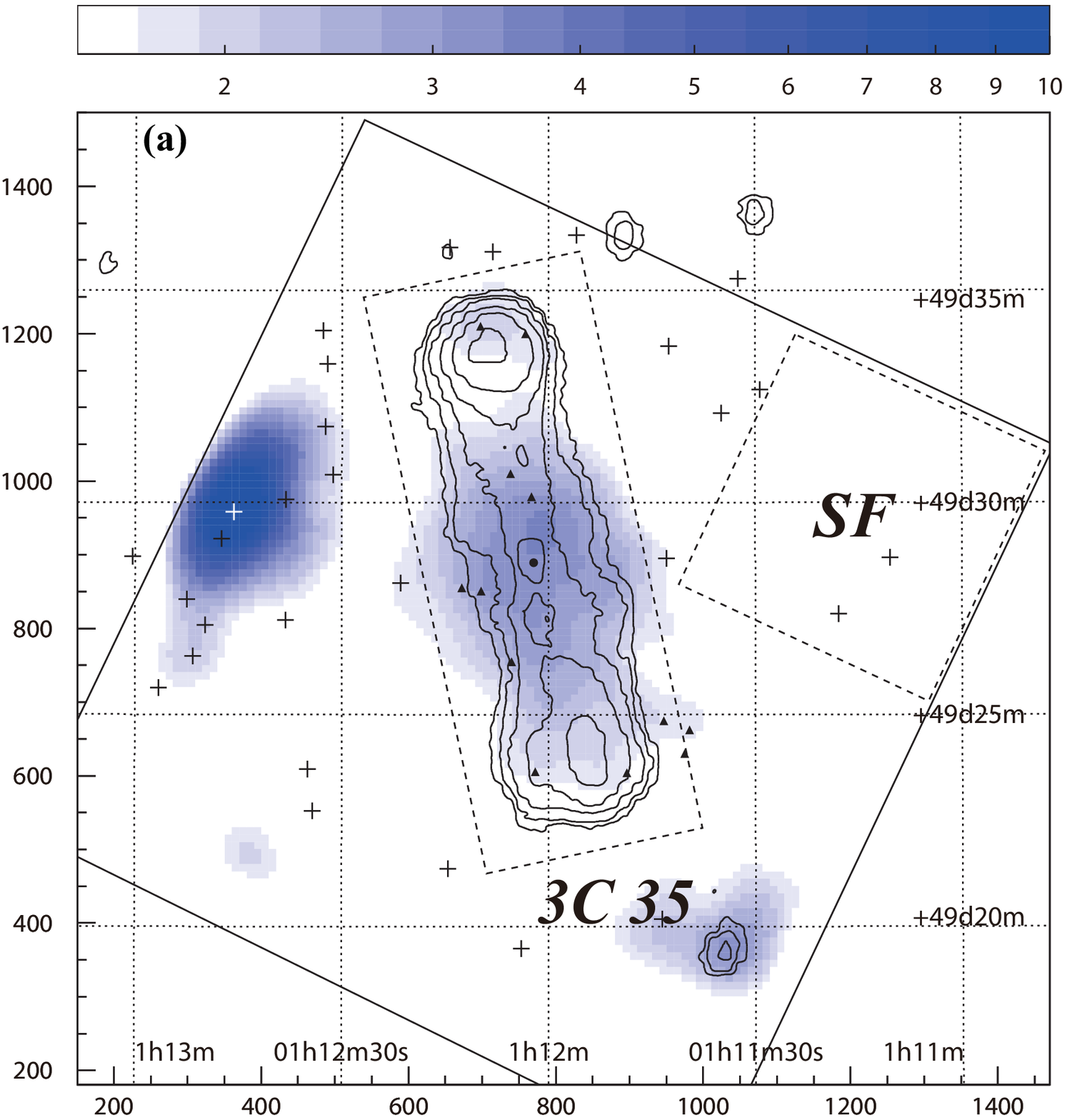}
\includegraphics[width=8.0cm]{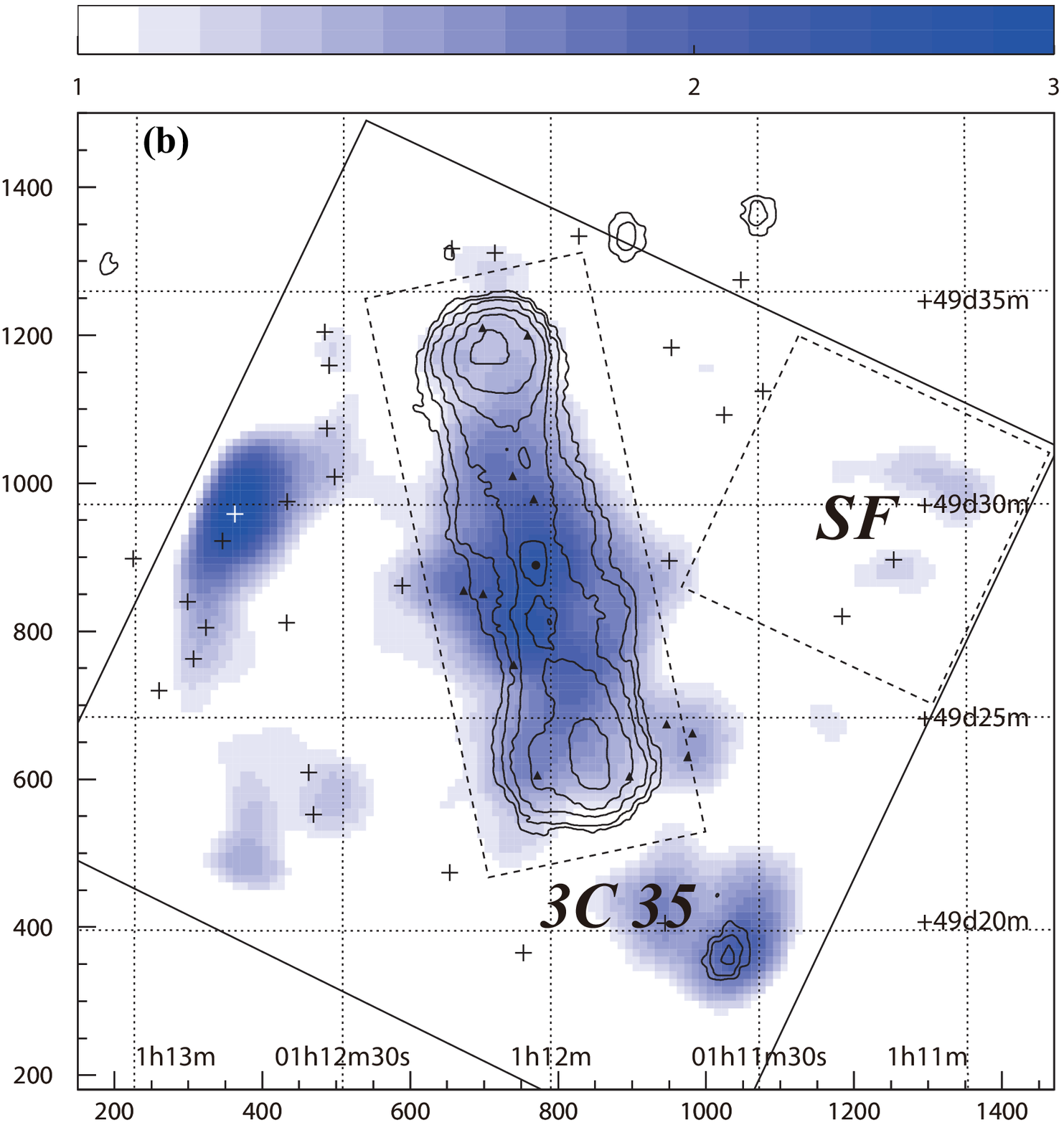}
\includegraphics[width=8.0cm]{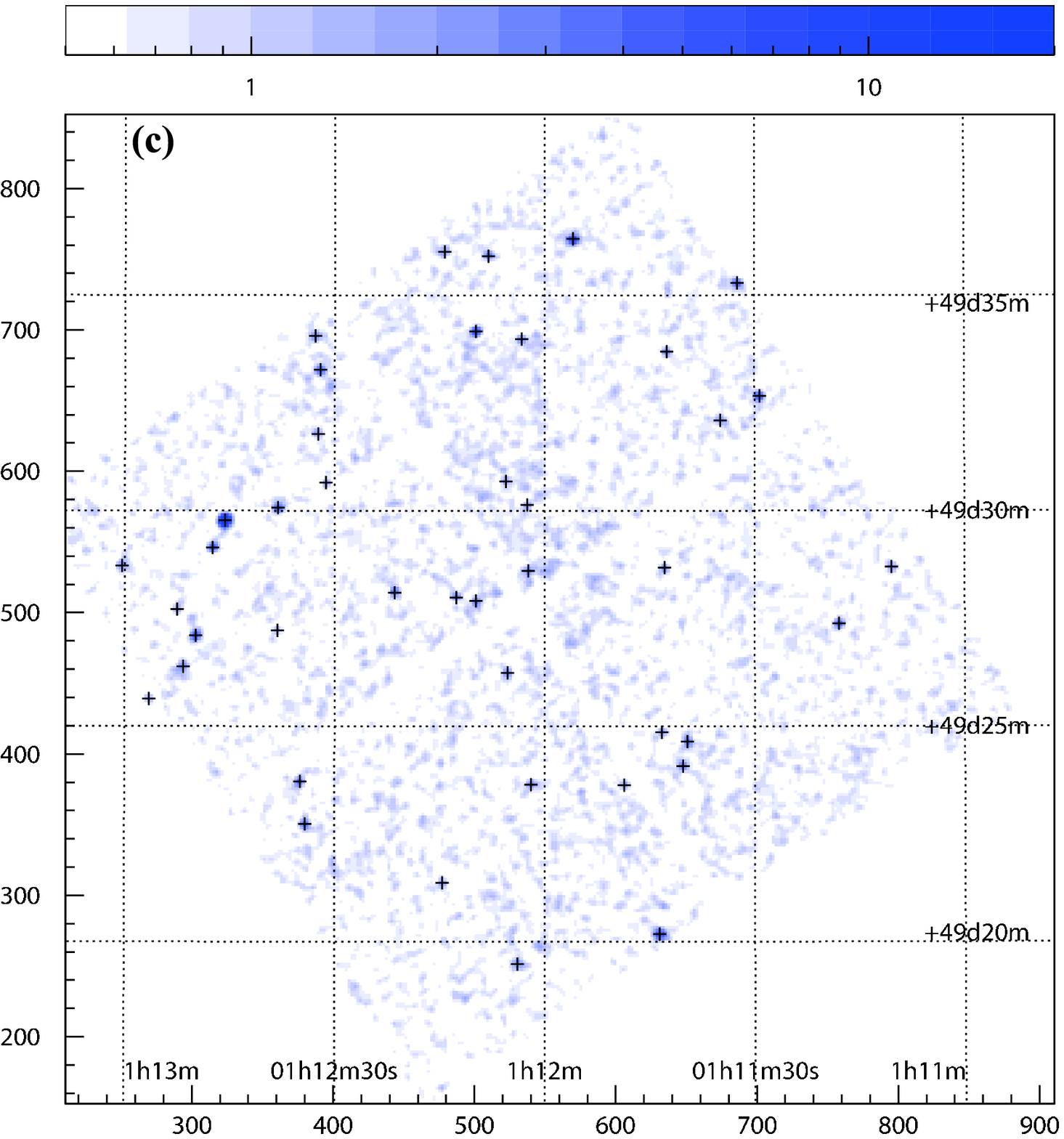}
\end{center}
\caption{The \suzaku~XIS image of 3C 35 
in the 0.5 -- 2 keV (panel a) and 2 -- 5.5 keV (panel b) ranges,
heavily smoothed with a two-dimensional Gaussian function of a $30''$ radius.
The scale bar indicates 
the X-ray counts integrated within a $10'' \times 10''$ bin.  
A 608.5 MHz radio contour image (Leahy et al. unpublished) is superposed.
The XIS FoV is shown with the solid square.
The sources detected with \chandra~are indicated with crosses or triangles.
Those with the triangles are taken into account in the spectral analysis.
The circle and white crosses point the positions 
of the nucleus of 3C 35 and \rxs, respectively. 
X-ray signals of 3C 35 and the source free region were
accumulated from the dashed rectangle and square,  
labeled as {\bf\itshape 3C 35} and {\bf\itshape SF}, respectively.
The 0.3 -- 10 keV \chandra~ACIS image of the same field is displayed 
in panel (c), after smoothing with a Gaussian of a $4''$ radius. 
The scale bar shows the X-ray counts in a $4'' \times 4''$ bin.  
}
\label{fig:image}
\end{figure*}

Located at the redshift of 
$z = 0.0670$ \citep{3C35_redshift},
3C 35 is a low-excitation radio galaxy 
\citep[LERG,][]{RG_summary,RG_summary2}
with an elliptical host \citep{elliptical_host}.  
The radio images of \src~revealed its classical Fanaroff-Riley (FR) II 
morphology \cite[e.g.,][]{3C35_RadioImage},
with a total angular size of $\sim 12'.5$,
corresponding to $\sim 950$ kpc at the source rest frame. 
The spectral age of \src~was estimated
as $\sim143$ Myr \citep{3C35_RadioIndex},
indicating that this radio galaxy is actually an evolved and old source.
A moderate radio intensity of the object, 
$2.3$ Jy at $1.4$ GHz \citep{radio_flux2}, 
in combination with its large physical size,  
ensures a high IC X-ray intensity. 
Thus, this giant radio galaxy was selected as 
a suitable target for a \suzaku~observation. 

In the present paper, the cosmological constants 
of $H_{\rm 0} = 71$ km s$^{-1}$ Mpc$^{-1}$, 
$\Omega_{\rm m} = 0.27$, and  $\Omega_{\lambda} = 0.73$
are adopted. 
This cosmology gives the luminosity distance of $ 297.8 $  Mpc
and the angle-to-size conversion ratio 
of $76.1$ kpc/$1'$, at the redshift of 3C 35.

%%%%%%%%%%%%%
% Section 2 %
%%%%%%%%%%%%%
\section{Observation and Data Reduction} %====================
\subsection{\suzaku~Observation} %====================
\label{sec:suzaku_obs}
The giant radio galaxy 3C 35 was observed 
with \suzaku~on 2010 January 4 -- 6. 
The XIS was operated in the normal clocking mode with no window option,
while the Hard X-ray Detector \citep{HXDdesign} was in the normal mode. 
In order to avoid the lobes of \src~falling 
on the anomalous columns in the Segment A of the XIS 0
\footnote{{\tt http://www.astro.isas.jaxa.jp/suzaku/news/2009/0702/}},
the XIS nominal position for the X-ray telescope \citep[XRT;][]{XRTpaper} 
was placed $1'.3$ south of the center of its host galaxy,
with the satellite roll angle fixed at $ 245^\circ$
\footnote{The definition of the roll angle of the \suzaku~satellite is 
described in \S 7 %X-Ray Imaging Spectrometer (XIS)
of ''The \suzaku~Technical Description'' ; 
{\tt http://heasarc.gsfc.nasa.gov/docs/suzaku/prop\_tools/suzaku\_td/}}. 
In this set up, the whole radio structure of 3C 35 
($\sim 12'.5$ in the north-south direction) was 
safely observed within the clean XIS field of view,  
% (a $17'.8 \times 17'.8$ square),
without any contamination from the calibration source 
at the specific corners of the CCD chips.

The data reduction and analysis were performed 
with the standard software package, HEASOFT 6.8. 
The X-ray emission from the source is very faint in the HXD bandpass, 
and only the XIS data are utilized in the present paper.
The cleaned event files created in the standard processing 
were directory utilized, without reprocessing. 
Correspondingly, 
the CALDB were utilized,
as of 2009 September 25 for the XIS and of 2008 July 09 for the XRT. 
According to the standard manner, 
the following criteria were adopted for the data screening; 
the spacecraft is outside the south Atlantic anomaly (SAA),
the time after an exit from the SAA is larger than 436 s,
the geometric cut-off rigidity is higher than $6$ GV, 
the source elevation above the rim of bright and night Earth is 
higher than $20^\circ$ and $5^\circ$, respectively, 
and the XIS data are free from telemetry saturation. 
These procedures yielded about $71$ ks of good exposure.  
In the scientific analysis below,
those events with a grade of 0, 2, 3, 4, or 6 were accumulated.

\subsection{Archival \chandra~data}%====================
On 2009 March 8 -- 9,
3C 35 was observed with the \chandra~ACIS (ObsID = 10240). 
In order to assist in the interpretation of the \suzaku~results,
the \chandra~data in this observation were analyzed. 
All the data were reprocessed in the standard manner
to create the new level 2 event file, 
with the CIAO 4.2 software package, referring to CALDB 4.2.1. 
The {\tt acis\_detect\_afterglow} correction was removed,
and a ``new'' bad pixel file created 
by the {\tt acis\_run\_hotpix} tool was applied. 
Because the background level was found to be 
stable during the observation, 
no additional data screening was performed 
to the new level 2 event file. 
As a result, 
the good exposure became $25.6$ ks for the observation.
In the following analysis, 
the grade selection same as for the \suzaku~XIS data 
(i.e., 0, 2, 3, 4, or 6) was adopted. 

%==========% 
% Figure 2 %
%==========%
\begin{figure}[t]
\begin{center}
\includegraphics[angle=-90,width=8cm]{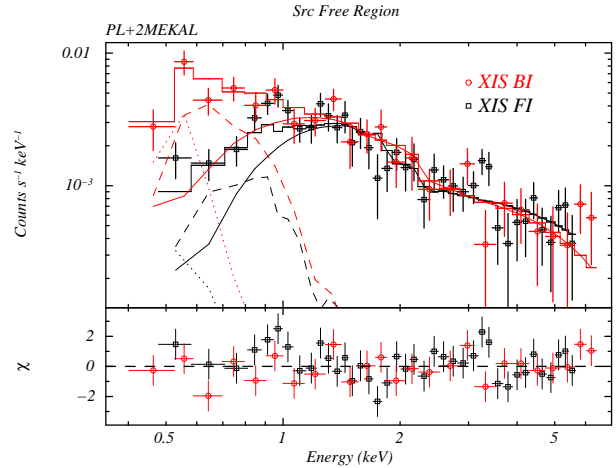}
\end{center}
\caption{NXB-subtracted XIS spectrum of the SF region, 
without removing the instrumental response.
The best-fit model,
consisting of a hard PL component with a photon index of 
$\Gamma = 1.41$ (solid line) 
and two MEKAL ones with the temperatures fixed at 
$kT_{\rm 1} = 0.204$ keV and $kT_{\rm 2} = 0.74$ keV 
(dotted and dashed lines, respectively),   
is indicated by histograms in the top panel,
while the residuals to the model are displayed in the bottom panel. }
\label{fig:SrcFree}
\end{figure}

%%%%%%%%%%%%%
% Section 3 %
%%%%%%%%%%%%%
\section{Results} %=====================================================
\label{sec:results}
\subsection{X-ray Image} %----------------------------------------------
\label{sec:image}
The background-inclusive \suzaku~XIS images of \src,
in the soft ($0.5$ -- $2$ keV) and hard ($2$ -- $5.5$ keV) bands,
are shown in the panels (a) and (b) of Figure \ref{fig:image}, 
respectively, 
after being heavily smoothed with a two-dimensional Gaussian kernel 
of a $30''$ radius.
On the XIS images, 
the 608.5 MHz radio contour map (Leahy et al. unpublished) 
\footnote{
Taken from ``An Atlas of DRAGNs'', edited by Leahy, Bridle, \& Strom;
{\tt http://www.jb.man.ac.uk/atlas/}.   }
are superposed. 
A systematic offset of $\sim 20''$ was noticed in the XIS coordinate, 
which may be regarded to be within the current systematic uncertainties 
in the XIS position determination \citep{Suzaku_pos_acc}. 
The XIS image was shifted, by referring to the position of 
a bright serendipitous X-ray source within the XIS field of view,
\rxs~\citep[][;the white cross in the XIS image]{1RXS},  
which was determined by the \chandra~ACIS 
to be located at (R.A., DEC) = ($18^\circ.1904$, $+49^\circ.4962$)
with an error of $\Delta\theta < 0.5''$. 

The XIS images suggest that the X-ray emission 
from the nucleus of \src~(the filled circles on the XIS images) was weak.
Instead, in these images,
faint and possibly diffuse X-ray emission,
apparently associated with \src, is clearly detected. 
In the soft X-ray image (panel a of Figure \ref{fig:image}),
the faint emission appears to be somewhat concentrated 
around the \src~nucleus or its host galaxy. 
From this spatial distribution,
the soft component is suggested to originate 
from the thermal plasma in the \src~host galaxy.
In contrast, 
the hard band X-ray image seems to be more elongated
along the radio lobes of \src,
as is seen in the panel (b) of Figure \ref{fig:image}. 
This indicates that the emission from the radio electrons in the lobes 
has a significant contribution to the hard band XIS image. 

The XIS image is thought to be inevitably contaminated 
by faint point-like X-ray sources, 
which were unresolved with the XIS spatial resolution 
\citep[a half power diameter of $\sim 2'$;][]{XRTpaper}.
In order to search for those contaminant X-ray sources, 
the $0.3$ -- $10$ keV \chandra~ACIS images 
of the same sky field is shown in the panel (c) 
of Figure \ref{fig:image}. 
Using the CIAO tool {\tt wavdetect}, 
42 X-ray sources were detected in the ACIS image,
including \rxs~mentioned above, 
and they are overlaid on the XIS images.
Because several sources appear to contaminate 
the diffuse X-ray emission detected with the XIS,  
the contribution from these sources 
should be considered adequately in the spectral analysis below.
It is consistent with the XIS result that 
the \src~nucleus was rather faint in the \chandra~observation, 
as is investigated in \S \ref{sec:chandra_src}. 

\subsection{X-ray Emission Associated with \src} %----------------------------
\label{sec:3C35}
In this section, the \suzaku~XIS spectrum of the faint X-ray emission
associated with the lobes and host galaxy of \src~is carefully examined. 
For this purpose, the impact of the background and 
contaminating point sources is taken into account step by step. 

\subsubsection{\suzaku~spectrum of the source free region}  %-------------
\label{sec:suzaku_SF}
In order to examine the X-ray spectrum of the diffuse emission 
with a low surface brightness, 
like that associated with the lobes and host galaxy of \src, 
it is of crucial importance to determine 
accurately the X-ray background (XRB) level.
Therefore, in the fist step, 
the XRB spectrum in this field was examined, referring to \citet{3C326}.
The XRB signal events were accumulated from the source free (SF) region,
which is shown with the dashed square denoted as {\bf\itshape SF} 
in Figure \ref{fig:image}.
This region was selected to 
avoid the XIS 0 anomalous columns located to the south of \src. 
The spectrum of the non-X-ray background (NXB) was estimated 
by the HEADAS tool {\tt xisnxbgen}.
The accuracy of the tool is reported to be better than 
$\sim 3$\% in the 1 -- 7 keV range, 
for a typical exposure of $50$ ks \citep{xisnxbgen}.
Figure \ref{fig:SrcFree} shows the NXB-subtracted XRB spectrum in this field.
The {\itshape SF} region is unfortunately irradiated 
with a radioactive calibration source ($^{55}$Fe)
for the front-illuminated (FI) CCD chip of the XIS 
\citep[XIS 0 and 3; ][]{XISpaper}. 
Therefore, the FI data are limited below $5.5$ keV,
while those from the backside-illuminated (BI) CCD chip (XIS 1) 
were utilized up to $\sim 6.5$ keV. 

%---------%
% Table 1 %
%---------%
\begin{table}[t]
\caption{Best-fit spectral parameters for the source free region.}
\label{table:SRC_Free}
\begin{center}
\begin{tabular}{ll}
\hline \hline %===================================================
Parameter                                 & Value \\
\hline %------------------------------------------------
$N_{\rm H}$ ($10^{21}$ \colden)             & $1.21$ \tablenotemark{c} \\
$\Gamma $                                  & $1.41$ \tablenotemark{d}  \\
$f_{\rm PL}$ (\SB) \tablenotemark{a}        & $(6.0 \pm 0.5) \times 10^{-8}$  \\
$kT_{\rm 1}$ (keV)                          & $0.204$ \tablenotemark{e}  \\
$kT_{\rm 2}$ (keV)                          & $0.074$ \tablenotemark{e}  \\
$f_{\rm th}$ (\SB) \tablenotemark{b}        & $(1.3 \pm 0.4) \times 10^{-8}$\\
$\chi^2/{\rm dof}$                         & $ 62.7 / 61$ \\
\hline %------------------------------------------------
\end{tabular} 
\end{center}
\tablenotetext{a}{The observed 2 -- 10 surface brightness of the PL component.}
\tablenotetext{b}{The observed 0.5 -- 2 keV surface brightness 
                  of the sum of the 2 MEKAL components.}
\tablenotetext{c}{Fixed at the Galactic value.  }
\tablenotetext{d}{Taken from \citet{asca_cxb}.  }
\tablenotetext{e}{Taken from \citet{xmm_cxb}. }
\end{table}

It is widely known that the XRB spectrum in the 0.2 -- 10 keV range 
is decomposed into a hard power-law (PL) component 
with a photon index of $\Gamma = 1.41$ \citep{asca_cxb}, 
and a two-temperature soft thermal plasma emission 
with temperatures of $kT_1 = 0.204$ keV and $kT_2 = 0.074$ keV \citep{xmm_cxb}.
The hard PL component,
which is reported to exhibit only a small spatial fluctuation
in its surface brightness \citep[$\lesssim 7$\%; ][]{asca_cxb},
is thought to be dominated by unresolved faint sources,
such as distant active galactic nuclei. 
In contrast, the soft thermal component, 
with a significant field-to-field intensity variation, 
is thought to be associated with the Galaxy.

The XIS spectrum of the XRB in the field was fitted with 
a composite model consisting of a PL component 
with the photon index fixed at $\Gamma = 1.41$ and 
of two soft MEKAL \cite[e.g.,][]{MEKAL} ones
with the temperatures fixed at $kT_1 = 0.204$ keV and $kT_2 = 0.074$ keV.
All the components were subjected to a photoelectric absorption
with the Galactic hydrogen column density in the direction of 3C 35 
\citep[$N_{\rm H} = 1.21 \times 10^{21}$ cm$^{-1}$; ][]{NH}.
Because the result was found to be insensitive 
to the metalicity of the MEKAL components,
the solar abundance ratio was adopted.
The tool {\tt xisrmfgen} was adopted to calculate   
response matrix functions (rmf) of the XIS.
Auxiliary response files (arf) were generated 
by {\tt xissimarfgen} \citep{xissimarf}, 
assuming a diffuse source 
with a $20'$ radius and a uniform surface brightness distribution. 

An acceptable fit ($\chi^2/{\rm dof} = 62.7 / 61$) was obtained 
by the model, with the parameters summarized in table \ref{table:SRC_Free}. 
The total absorption-inclusive surface brightness of the XRB was measured as 
$f = (8.9 \pm 0.3) \times 10^{-8}$ \SB, in the 0.5 -- 10 keV range.
It is important to note that 
the 2 -- 10 keV surface brightness of the hard PL component,  
$f_{\rm PL} = (6.0 \pm 0.5) \times 10^{-8}$ \SB, 
was found to be perfectly consistent with the result in \citet{asca_cxb}, 
$f_{\rm PL} = (6.4 \pm 0.6) \times 10^{-8}$ \SB.
Therefore, this result ensures that 
the best-fit XRB model is safely adopted for the spectrum of \src,
in combination with the NXB spectrum reproduced by {\tt xisnxbgen}.

\subsubsection{\chandra~spectrum of contaminating sources}  %-----------------
\label{sec:chandra_src}
In the next step, the contamination from the X-ray point sources 
detected with \chandra~is evaluated.
The low signal statistics with the \chandra~ACIS prevented us 
from analyzing the X-ray spectra of the individual sources. 
Therefore, the 13 sources,
which are thought to contaminate significantly, are selected,
and the sum of their spectra was analyzed.
These sources are shown with the triangles and circle (the \src~nucleus)
in Figure \ref{fig:image}, 
and their positions and ACIS count rates are tabulated 
in Table \ref{table:contamination}. 
The contribution from the other sources 
to the XIS spectrum of \src~discussed in \S \ref{sec:suzaku_3c35}
is estimated to be less than $10$ \% of the sum flux of the 13 sources,
and so is regarded to be negligible.

%---------%
% Table 2 %
%---------%
\begin{table}[t]
\caption{Contaminating point sources  
taken into account in the spectral analysis.}
\label{table:contamination}
\begin{center}
\begin{tabular}{llc}
\hline  \hline %===================================================
(R.A., Dec)        & $\Delta\theta$\tablenotemark{a} & Signal\tablenotemark{b} \\  
\hline %---------------------------------------------------------
($ 17.9529 $, $ +49.3939 $ ) & $1.0$ & $ 5.8 \pm 1.8 $  \\ % 4 
($ 18.0082 $, $ +49.3942 $ ) & $0.7$ & $ 7.0 \pm 2.0 $  \\ % 5 
($ 17.9177 $, $ +49.4013 $ ) & $0.6$ & $ 9.7 \pm 2.2 $  \\ % 7 
($ 17.9150 $, $ +49.4107 $ ) & $0.8$ & $ 11.5 \pm 2.5 $ \\ % 8 
($ 17.9303 $, $ +49.4142 $ ) & $1.1$ & $ 6.8 \pm 2.0 $  \\ % 9 
($ 18.0222 $, $ +49.4374 $ ) & $0.9$ & $ 9.3 \pm 2.2 $  \\ % 10 
($ 18.0409 $, $ +49.4652 $ ) & $0.8$ & $ 7.0 \pm 2.0 $  \\ % 16 
($ 18.0525 $, $ +49.4665 $ ) & $0.6$ & $ 7.7 \pm 2.0 $  \\ % 17 
($ 18.0096 $, $ +49.4768 $ ) \tablenotemark{c}
                             & $0.7$ & $ 6.0 \pm 1.9 $  \\ % 19 
($ 18.0104 $, $ +49.5024 $ ) & $0.7$ & $ 5.3 \pm 1.7 $  \\ % 24 
($ 18.0230 $, $ +49.5114 $ ) & $0.8$ & $ 4.6 \pm 1.7 $  \\ % 26 
($ 18.0138 $, $ +49.5664 $ ) & $0.5$ & $ 6.5 \pm 1.8 $  \\ % 32 
($ 18.0410 $, $ +49.5693 $ ) & $0.3$ & $ 21.5 \pm 3.2 $ \\ % 34 
\hline %------------------------------------------------
\end{tabular}
\end{center}
\tablenotetext{a}{The position error in arcsec}
\tablenotetext{b}{The 0.3 -- 10 keV count rate of the source 
                  in the unit of $10^{-4}$ \rate }
\tablenotetext{c}{Corresponding to the nucleus of \src.}
\end{table}

The ACIS events were integrated within a circle with a $9.8''$ radius 
centered on each source,  
while the background ones were derived from neighboring source free regions 
each with a $19.7''$ radius. 
The rmf and arf were generated the CIAO tools 
{\tt mkacisrmf} and {\tt mkarf}, respectively,
for the individual sources, and they are simply averaged.  
Figure \ref{table:spec_contami} shows the summed ACIS spectrum 
of the 13 sources in the $0.5$ -- $4$ keV range. 
It is described with a single PL model with parameters listed 
in Table \ref{table:spec_contami}.
The absorption column density is found to be consistent 
with the Galactic value within the statistical errors,
while the photon index was similar to the typical value 
for active galactic nuclei.
Thus, the result appears to be consistent with a picture that
these sources are dominated by distant active galaxies. 
The summed flux in 0.5 -- 5 keV 
was derived as $7.6_{-1.5}^{+0.7} \times 10^{-14}$ \flux,
without removing the absorption. 

Adopting this best fit model,
the count rate of the \src~nucleus in 0.3 -- 10 keV 
shown in Table \ref{table:contamination} was converted to 
the 2 -- 10 keV X-ray luminosity of $\sim 5 \times 10^{40}$ \lum. 
This value is at the lower end of the X-ray luminosities 
of typical LERGs \citep{RG_summary3}. 
Even if a heavy obscuration of $N_{\rm H} = 1 \times 10^{23}$ \colden,
like in the case of narrow line radio galaxies \citep{RG_summary3}, 
is assumed,
the absorption-corrected X-ray luminosity is estimated as 
at most $\sim 4 \times 10^{41}$ \lum~in the 2 -- 10 keV range.
Thus, the \src~nucleus is suggested to be relatively inactive,
or radiatively inefficient. % \citep{RG_summary3}.

%==========% 
% Figure 3 %
%==========%
\begin{figure}[t]
\begin{center}
\includegraphics[angle=-90,width=8cm]{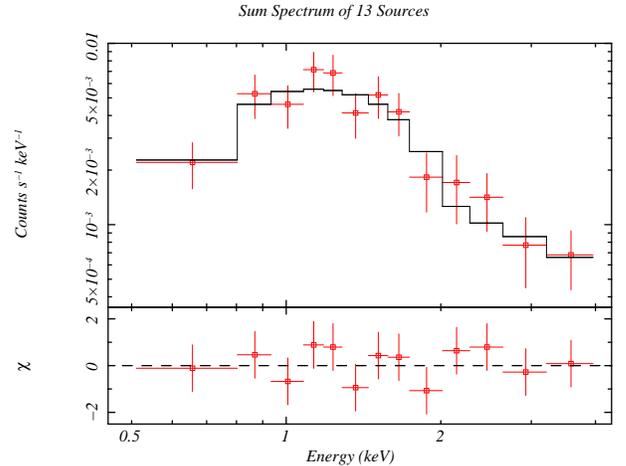}
\end{center}
\caption{Sum of the 0.5 -- 4 keV \chandra~ACIS spectra from the 13 sources, 
which can contaminate the \suzaku~spectrum of the \src~region.
The best-fit PL model is shown with the histogram.}
\label{fig:contamination}
\end{figure}

%---------%
% Table 3 %
%---------%
\begin{table}[t]
\caption{Best-fit parameters 
for the summed spectrum of the 13 contaminating sources }
\label{table:spec_contami}
\begin{center}
\begin{tabular}{ll}
\hline  \hline %===================================================
Parameters     &  Value    \\
\hline %------------------------------------------------
$N_{\rm H}$  ($10^{21}$ \colden)  & $0.44 (<2.74)$          \\
$\Gamma$                        & $1.86_{-0.39}^{+0.80}$          \\
$F_{\rm X}$  ($10^{-14}$ \flux) \tablenotemark{a}  
                                & $7.6_{-1.5}^{+0.7} $          \\
$\chi^2/{\rm d.o.f}$            & $5.6/10$           \\
\hline %------------------------------------------------
\end{tabular}
\end{center}
\tablenotetext{a}{The observed X-ray flux in 0.5 -- 10 keV.}
\end{table}

\subsubsection{\suzaku~spectrum of \src}  %------------------------------------
\begin{figure*}[t]
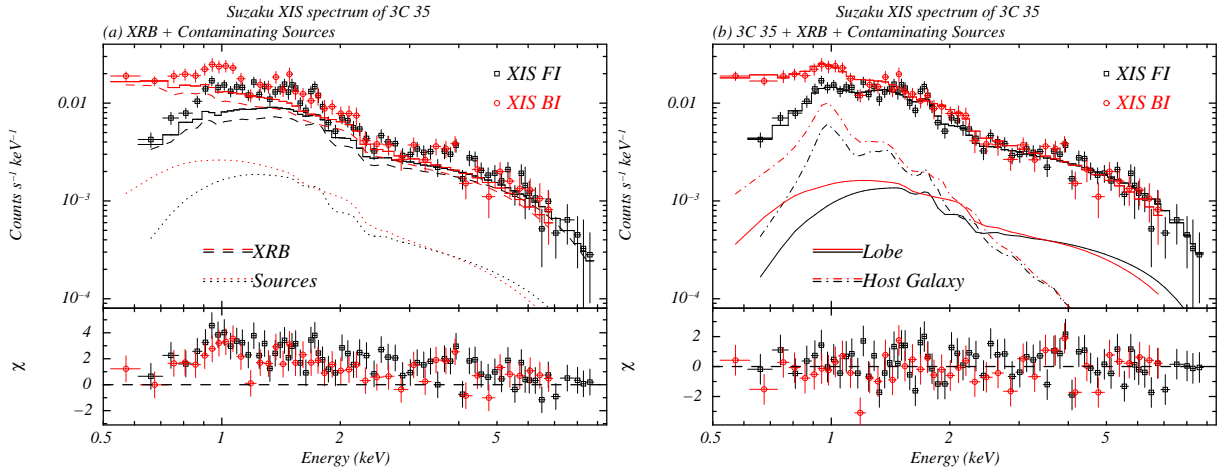

\begin{center}
\includegraphics[angle=-90,width=8cm]{figure4a.ps}
\includegraphics[angle=-90,width=8cm]{figure4b.ps}
\end{center}
\caption{The XIS spectrum of the \src~region. 
The panel (a) compares the data with the model,
describing the XRB (the dashed lines) 
and contaminating sources (the dotted lines),
derived in \S \ref{sec:suzaku_SF}  and \S \ref{sec:chandra_src}, respectively. 
The panel (b) shows the best-fit model,
after the contributions from the host galaxy (MEKAL; the dash-dotted lines) 
and the lobe (PL; the solid line) are considered. 
The XRB and contaminating source components are omitted in the panel (b), 
for clarity.  
}
\label{fig:3C35}
\end{figure*}

\label{sec:suzaku_3c35}
% Signal Statistics : See the note 2 page 15 % 
Finally, the \suzaku~XIS spectrum of the faint diffuse X-ray emission, 
associated with \src~was integrated within the rectangle, 
denoted as {\bf\itshape 3C35} in Figure \ref{fig:image}.
This region was carefully determined 
to include the whole radio structure of \src.
Figure \ref{fig:3C35} shows the XIS spectrum of \src,
after subtracting the NXB spectrum in the region 
simulated by the {\tt xisnxbgen} tool. 
The data and NXB count rates in the 0.7 -- 7 keV range were measured  
as $(3.58 \pm 0.05) \times 10^{-2}$ \rate~and 
$(0.90 \pm 0.02) \times 10^{-2}$ \rate, respectively,  per one FI CCD chip,
while those with the BI chip were derived 
as $(4.93 \pm 0.08) \times 10^{-2} $ \rate~and
$(1.71 \pm 0.04) \times 10^{-2} $ \rate.
Here, only the statistical errors are shown.
Thus, a statistically significant signal was detected above the NXB,
with the 0.7 -- 7 keV FI and BI count rates of 
$(2.69 \pm 0.05) \times 10^{-2} $ \rate~and 
$(3.22 \pm 0.09) \times 10^{-2}$ \rate, respectively.
The signal count rate is considerably larger than 
the typical NXB uncertainty of 
$\sim 3$\% \citep{xisnxbgen}, corresponding to the FI and BI count rates of 
$0.03 \times 10^{-2} $ \rate~and $0.05 \times 10^{-2}$ \rate. 

%==========% 
% Figure 4 %
%==========%

%---------%
% Table 4 %
%---------%
\begin{table}[t]
\caption{Summary of the PL+MEKAL fitting to the excess signals from \src}
\label{table:3C35}
\begin{center}
\begin{tabular}{lllllll}
\hline \hline %===================================================
Parameters                          & \multicolumn{3}{l}{Case 1} &\multicolumn{3}{l}{Case 2} \\
\hline %------------------------------------------------
$N_{\rm H}$ ($10^{21}$ cm$^{-2}$)   & \multicolumn{3}{l}{1.21\tablenotemark{b}}  &\multicolumn{3}{l}{1.21\tablenotemark{b}} \\
$\Gamma $                           & $1.35$ & $_{-0.86}^{+0.56}$ & $_{-0.10}^{+0.11}$           &  1.7 \tablenotemark{c} &    &      \\
$S_{\rm 1 keV}$ (nJy)               & $8.6 $ & $_{-6.9}^{+10.6}$  & $_{-2.4}^{+3.2}$             & $13.6$  &$ \pm 5.4 $        & $_{-3.6}^{+4.0}$     \\
$kT$ (keV)                          & $1.36$ & $_{-0.23}^{+0.28}$ & $\pm0.02$                    & $1.33$  &$_{-0.21}^{+0.27}$ & $\pm0.03$     \\
$L_{\rm X}$ ($10^{41}$ \lum)\tablenotemark{d}         
                                    & $9.0$  & $_{-4.9}^{+5.9}$   & $_{-0.1}^{+0.2}$             & $7.5$   &$_{-3.7}^{+2.8} $  & $\pm0.6$     \\
$\chi^2/{\rm dof}$                  & \multicolumn{3}{l}{$105.1/108$}                            & \multicolumn{3}{l}{$106.0/109$}   \\
\hline %------------------------------------------------
\end{tabular} 
\tablenotetext{a}{The first and second errors represent the statistical and systematic ones, respectively.}
\tablenotetext{b}{Fixed at the Galactic Value.}
\tablenotetext{c}{Fixed at the radio synchrotron index.}
\tablenotetext{d}{Absorption-corrected 0.5 -- 10 keV luminosity of the MEKAL component.}
\end{center}
\end{table}

In the panel (a) of Figure \ref{fig:3C35}, 
the NXB-subtracted XIS spectrum of \src~is 
compared with the contributions from the XRB and contaminating sources. 
The best-fit model to the XRB spectrum determined in \S \ref{sec:suzaku_SF},
was convolved with the arf and rmf of the \src~region, 
which were calculated in the similar manner to the SF region. 
The $0.7$ -- $7$ keV FI and BI count rates of the XRB were 
estimated as $(1.57\pm0.06) \times  10^{-2}$ \rate~
and $(1.99\pm0.07) \times  10^{-2}$ \rate, 
respectively.
For the individual sources considered in \S \ref{sec:chandra_src},
the arfs were calculated, 
and averaged according to the \chandra~ACIS count rates. 
The best-fit \chandra~model to the sum spectrum of the contaminating sources 
are convolved with the averaged arf and rmf,
to derive the FI and BI count rates of 
$ 0.31_{-0.06}^{+0.03} \times  10^{-2}$ \rate~and 
$ 0.39_{-0.08}^{+0.03} \times  10^{-2}$ \rate, respectively. 
As a result, significant excess has remained 
with the FI and BI count rates in $0.7$ -- $7$ keV of 
$(0.80 \pm 0.05 _{-0.07}^{+0.09} ) \times 10^{-2}$ \rate~and 
$(0.85 \pm 0.09 _{-0.09}^{+0.12} ) \times 10^{-2}$ \rate, respectively, 
over the contribution from the XRB and contaminating sources.
This excess is clearly seen in the residual spectrum 
in the panel (a) of Figure \ref{fig:3C35}.  
Here and hereafter, 
the first error is due to the signal statistics of the \src~region,
while the errors from all the other components, 
including those from the NXB, XRB, contaminating sources and so forth, 
were propagated to the second one. 

In order to reproduce the excess X-ray spectrum from \src,
a composite model consisting of PL and MEKAL components,
both of which were subjected to the Galactic absorption 
($N_{\rm H} = 1.21 \times 10^{21}$ cm$^{-1}$), was utilized. 
Based on the XIS image in Figure \ref{fig:image}, 
the PL component was introduced to describe  
the emission to be associated with the lobes of \src,
while the MEKAL component corresponds to the thermal plasma emission 
from the host galaxy.
The metal abundance in the MEKAL model was fixed at 0.3 times the solar value,
the typical value for nearby elliptical galaxies 
\citep[e.g.,][]{Abund_elliptical}. 
The arf for the PL component was created by assuming 
a uniform rectangular emission region with size of $12'.5 \times 5'.0$,
based on the radio image \citep{3C35_RadioIndex}.
A point-like X-ray source at the center of the host galaxy was 
simply adopted, for the calculation of the MEKAL arf.
Here, the lobe arf is estimated to be smaller then the point source one,
by $20$ -- $30$ \% in 0.7 - 7 keV.  
As revealed by the \chandra~data, the nucleus of \src~was so faint that 
it was already taken into account as one of the contaminating sources. 

As shown in the panel (b) of Figure \ref{fig:3C35},
the PL+MEKAL model is found to reproduce the excess spectrum 
($\chi^2/{\rm d.o.f} = 105.1/108$)
with the parameters tabulated in Table \ref{table:3C35} (Case 1).
However, the errors from the model appear to be considerably large, 
since the two spectral components were strongly coupled to each other. 
The best-fit photon index of the PL component, 
$\Gamma = 1.35_{-0.86}^{+0.56}$$_{-0.10}^{+0.11}$,
was found to be 
consistent with the synchrotron radio photon index from the lobe 
\citep[$\Gamma_{\rm R} = 1.7$ between $73.8$ -- $327.4$ MHz;][]
{3C35_RadioIndex}, 
and was fixed at this value.
The model again became acceptable ($\chi^2/{\rm d.o.f} = 106.0/109$), 
giving the best-fit parameters in Table \ref{table:3C35} (Case 2). 
The flux density at 1 keV of the PL component was derived 
as $S_{\rm 1 keV} = 13.6\pm 5.4 _{-3.6}^{+4.0}$ nJy. 
The MEKAL temperature was measured as $kT = 1.33_{-0.21}^{+0.27} \pm 0.03$ keV,
while its absorption-corrected 0.5 -- 10 keV luminosity was evaluated as 
$L_{\rm X} = (7.5 _{-3.7}^{+2.8} \pm 0.6) \times 10^{41}$ \lum, 
at the redshift of 3C 35 ($z = 0.0670$).

\section{Discussion} %-----------------------------------
\begin{figure}[t]
\begin{center}
\plotone{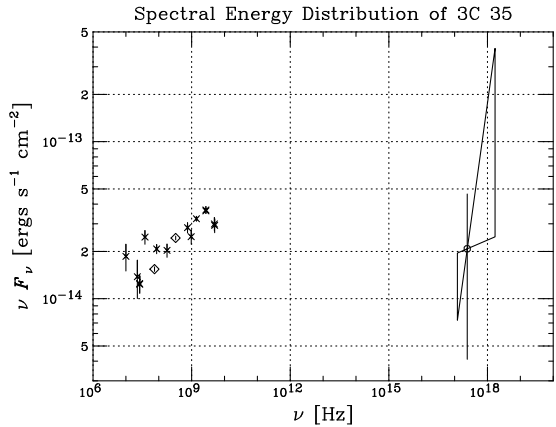}
\end{center}
\caption{Spectral energy distribution of 3C 35.
The best-fit PL model to the X-ray spectrum with a free photon index 
is plotted by the bow tie. 
The diamonds indicate the recent radio data from \citet{3C35_RadioIndex},
which are utilized to evaluate the physical parameters,
while the crosses represent those from \citet{radio_flux1,radio_flux2}.
}
\label{fig:sed}
\end{figure}

The \suzaku~XIS image shown in Figure \ref{fig:image}
revealed faint extended X-ray emission associated with 
the host galaxy and radio lobes of \src.
After carefully subtracting the NXB, XRB, and 
the contamination from faint X-ray sources detected with \chandra, 
the \suzaku~XIS spectrum from \src~was decomposed 
into the PL and MEKAL components,
as is displayed in Figure \ref{fig:3C35}. 
The PL component is found to dominate the MEKAL one above $\sim 2$ keV. 
In combination with the XIS images (Figure \ref{fig:image}), 
in which the X-ray emission appears to be more extended along the radio lobes
in the hard ($2$ -- $5.5$ keV) X-ray band, 
the PL component is regarded as emission from the lobes. 
The spectral energy distribution of the synchrotron radio emission 
and the X-ray PL component from \src~is shown in Figure \ref{fig:sed}. 
The agreement between the X-ray photon index 
$\Gamma = 1.35_{-0.86}^{+0.56}$$_{-0.10}^{+0.11}$ 
and the radio synchrotron one 
\citep[$\Gamma_{\rm R} = 1.7$;][]{3C35_RadioIndex} 
supports an interpretation that the X-ray PL component 
is attributed to the IC emission from the electrons in the lobes,
which are also radiating the synchrotron radio photons. 
Based on the flux density at 1 keV of 
$S_{\rm 1 keV} = 13.6\pm 5.4 _{-3.6}^{+4.0}$ nJy,
derived when the photon index was fixed at $\Gamma_{\rm R} = 1.7$, 
and the large angular size ($\sim 12'.5 \times 5'$), 
the lobe X-ray surface brightness is measured as 
$\sim 0.2$ nJy arcmin$^{-2}$, 
which is amongst the faintest to be studied through IC X-ray emission 
to date \citep[e.g.,][]{3C326}. 

%==========% 
% Figure 5 %
%==========%

In contrast, the XIS image in the soft band ($0.5$ -- $2$ keV)
suggests that  the origin of the MEKAL component is a thermal plasma emission 
from the host galaxy,  
although an interpretation that it is originated 
from the possible group of galaxies surrounding the host galaxy 
is not rejected from the XIS data alone. 
The \chandra~data could potentially help to resolve the diffuse emission,
whose spatial extent is smaller than the XIS point spread function. 
Since the nucleus of 3C 35 is located close the gap of the ACIS chip
(see the panel (c) of Figure \ref{fig:image}), 
the analysis of the \chandra~data immediately around 3C 35 
is regarded to be difficult, and beyond the scope of the present paper.
The X-ray luminosity of the MEKAL component,
$L_{\rm X} = (7.5 _{-3.7}^{+2.8} \pm 0.6) \times 10^{41}$ \lum, 
derived with the PL photon index of $\Gamma_{\rm R} = 1.7$,
is found to be lower by a factor of $\sim 10$ than those anticipated 
for its temperature, $kT = 1.33_{-0.21}^{+0.27} \pm 0.03$ keV, 
from the luminosity-temperature relation of nearby clusters/groups 
of galaxies \citep[e.g,][]{cluster_LX-KT_2,cluster_LX-KT,cluster_LX-KT_3}. 
These two quantities are instead rather consistent 
with the luminosity-temperature relation of nearby elliptical galaxies 
\citep[e.g.,][]{Abund_elliptical,cluster_LX-KT}.
In addition, 
\src~was not suggested to be in a rich cluster environment \citep{RG-Cl}.
Thus, the host galaxy interpretation is thought to be more preferable 
for the MEKAL component.
It is important to note that 
any contamination from this component to the PL one is insignificant,
as long as the photon index is fixed at $\Gamma_{\rm R} = 1.7$.

%==========% 
% Figure 6 %
%==========%
\begin{figure}[t]
\begin{center}
\plotone{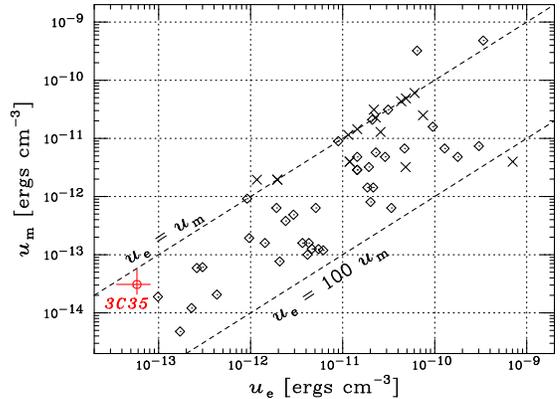}
\end{center}
\caption{
The relation between $u_{\rm e}$ and $u_{\rm m}$, in lobes of radio galaxies.
Those determined through the IC technique are plotted as diamonds 
\citep[][and reference therein]{lobes_Croston,3C326},
The red circle shows the results of 3C 35, 
where only the statistical error from $S_{\rm 1 keV}$ is plotted.
Lobes, from which only the upper limit of IC X-ray flux was obtained, 
are indicated by crosses.
The two dashed lines represents the equipartition and a particle dominance 
of $u_{\rm e} / u_{\rm m} = 1 $, and $100$, respectively.}
\label{fig:ue2um}
\end{figure}

%---------%
% Table 4 %
%---------%
\begin{table*}[t]
\caption{Physical parameters in the lobes of \src}
\label{table:energetics}
\begin{center}
\begin{tabular}{lll}
\hline \hline %===================================================
Parameters & Value & Comments \\
\hline %------------------------------------------------
$S_{\rm 1 keV}$ (nJy)    &  $13.6 \pm 5.4 _{-3.6}^{+4.0}$  & Case 2 in Table \ref{table:3C35}         \\
$S_{\rm R}$ (Jy)        &  $7.5 \pm 0.2 $                &  at $327.4$ MHz \\
$\Gamma_{\rm R}$           &   $1.70\pm0.03 $              &  between $73.8$ MHz and $327.4$ MHz \\
$V$ ($10^{72}$ cm$^{3}$) &  $3.2 \pm 0.3$                &          \\
\hline %------------------------------------------------
$u_{\rm e}$ ($10^{-14}$ \edens) \tablenotemark{a} &  $5.8 \pm 2.3 _{-1.7}^{+1.9}$      & $\gamma_{\rm e} = 10^{3}$ -- $10^{5}$\\
                                                 &  $16.2 \pm 6.4 _{-4.6}^{+5.2}$     & $\gamma_{\rm e} = 10^{2}$  -- $10^{5}$\\
$u_{\rm e}$ ($10^{-14}$ \edens) \tablenotemark{a} &  $3.1_{-1.0}^{+2.5}$$_{-0.9}^{+1.4}$    & \\
$B$ ($\mu$G) \tablenotemark{a}                 &  $0.88_{-0.16}^{+0.31}$$_{-0.14}^{+0.19}$ & \\
$u_{\rm e}/ u_{\rm m}$ \tablenotemark{a}         &  $1.9_{-1.3}^{+2.0}$$_{-1.0}^{+1.5}$ & $\gamma_{\rm e} = 10^{3}$ -- $10^{5}$\\
                                               &  $5.3_{-3.5}^{+5.6}$$_{-2.7}^{+4.1}$ & $\gamma_{\rm e} = 10^{2}$  -- $10^{5}$\\
\hline %------------------------------------------------
\end{tabular} 
\tablenotetext{a}{The first error is due to the statistical error of $S_{\rm 1 keV}$, 
                  while all the possible systematics are taken into account in the second error. }
\end{center}
\end{table*}

There are several seed photon candidates for the IC X-rays 
in lobes of radio galaxies, 
including the CMB radiation \citep{CMB_IC},
infrared (IR) photons from the nucleus \citep[e.g.,][]{IC_nuclearIR}, 
and the synchrotron radiation from the lobes themselves. 
At the redshift of \src~($z = 0.0670$), the CMB energy density is calculated 
as $u_{\rm CMB} = 5.3 \times 10^{-13}$ \edens. 
From the upper limit on the IR flux of the \src~nucleus 
of $20$ mJy at 12 $\mu$m \citep{IR_nucleus}, 
the nuclear IR photons are estimated to have an energy density 
of $ < 1.5 \times 10^{-15} (r/100{\rm~kpc})^{-2} $ \edens,
where $r$ is the distance from the nucleus,
assuming the emission from the nucleus is isotropic.
Except for the central region ($r \ll 10$ kpc), 
the nuclear IR photons are found to be unimportant. 
The synchrotron photons are fully negligible, 
since the radio flux density of 
\src~\citep[$ 7.5 \pm 0.2 $ Jy at $327.4$ MHz;][]{3C35_RadioIndex} 
gives a synchrotron photon energy density 
of $ \ll 10^{-17}$ \edens~averaged over the lobes.
Therefore, the CMB radiation is specified as the dominant seed photon source. 

In order to diagnose the energetics in the lobes of \src,
the synchrotron radio photon index and flux density were 
derived from the recent result presented in \citet{3C35_RadioIndex},
as $\Gamma_{\rm R} = 1.70 \pm 0.03$ and 
$S_{\rm R} = 7.5 \pm 0.2$ Jy at $324.7$ MHz, respectively. 
Correspondingly, the 1 keV IC X-ray flux density 
of $S_{\rm 1 keV} = 13.6\pm 5.4 _{-3.6}^{+4.0}$ nJy,
determined with the photon index fixed at $\Gamma_{\rm R}$ 
(Case 2 in Table \ref{table:3C35}), was utilized.
The number density spectrum of the radiating electrons 
is assumed to be a simple PL form 
of $\propto \gamma_{\rm e}^{2\Gamma_{\rm R} - 1}$,
where $\gamma_{\rm e}$ is the electron Lorentz factor. 
From the radio image \citep{3C35_RadioIndex}, 
the shape of the lobe is assumed to be a cylinder, 
with a radius and height of $190 \pm 10$ kpc and $950 \pm 10$ kpc, 
respectively. 
These give the total volume of the lobes 
as $(3.2 \pm 0.3)\times 10^{72}$ cm$^{3}$.

Table \ref{table:energetics} summarizes the energetics in the lobes,
evaluated by referring to \citet{CMB_IC},
together with the input observables discussed above.
The magnetic energy density spatially-averaged over the lobes 
of \src~was derived as 
$u_{\rm m} = (3.1_{-1.0}^{+2.5}$$_{-0.9}^{+1.4}) \times 10^{-14}$ \edens,
while the electron energy density with the rest mass energy subtracted 
was estimated as 
$u_{\rm e} = (5.8 \pm 2.3 _{-1.7}^{+1.9}) \times 10^{-14}$ \edens~ 
for $\gamma_e=10^3$ -- $10^5$, corresponding to the electrons 
observable directly through the synchrotron radio or IC X-ray emission. 
The first error is from the statistical one in $S_{\rm 1keV}$,
while in the second error, all the systematic ones from 
$S_{\rm 1keV}$, $S_{\rm R}$, $\alpha$ (or $\Gamma_{\rm R}$), 
and $V$ are considered.
Thus, an approximate equipartition condition 
of $u_{\rm e}/ u_{\rm m} = 1.9_{-1.3}^{+2.0}$$_{-1.0}^{+1.5}$
is realized in the lobes of \src.
As a result, the equivalent magnetic field strength of 
$B = 0.88_{-0.16}^{+0.31}$$_{-0.14}^{+0.19}$ $\mu$G
is consistent with the equipartition field 
\citep[$B_{\rm eq} \lesssim 0.9 $ $\mu$G;][]{3C35_RadioIndex}. 
Even if the minimum Lorentz factor was lowered down to $\gamma_e = 10^{2}$,
by referring to \cite{3C35_RadioIndex},
the energetics in the \src~lobes are found not to deviate 
significantly from equipartition, 
as $u_{\rm e}/u_{\rm m} = 5.3_{-3.6}^{+5.5}$$_{-2.7}^{+4.1}$,
though with the large errors.

Figure \ref{fig:ue2um} summarizes the relation 
between \ue~and \um~in the lobes of radio galaxies,
which were determined through the IC X-ray technique 
\citep[][and reference therein]{lobes_Croston,3C326}.
This diagram indicates that 
an electron dominance of $u_{\rm e}/u_{\rm m} \sim 10$ 
appears to be typical for the radio galaxies, 
over $\sim 4$ orders of magnitude in \ue~and \um. 
The \suzaku~result indicates that 
the $u_{\rm e}/u_{\rm m}$ ratio in the lobes of \src~
is lower than the typical value,
and even consistent with the equipartition,  
although with relatively large errors. 
In addition, \ue~and \um~in \src~are both located 
at the lower end of those of the radio galaxies
(the lower-left corner in Figure \ref{fig:ue2um}).
These features of \src~are thought to be related to the fact that 
it is an old radio source evolved to a size of $\sim 1$ Mpc,
with the spectral age of $143$ Myr \citep{3C35_RadioIndex}. 
Especially, \um~is found to be lower by an order of magnitude 
than the CMB energy density at the rest frame of \src.
This result confirms the dominance of IC energy losses over synchrotron ones,
which was predicted to be a common property of the giant radio galaxies 
under the equipartition assumption \citep{GRG_ICdominance}.

\begin{figure}[h]
\begin{center}
%\plotone{../fig/energetics/size2ue.ps}
\plotone{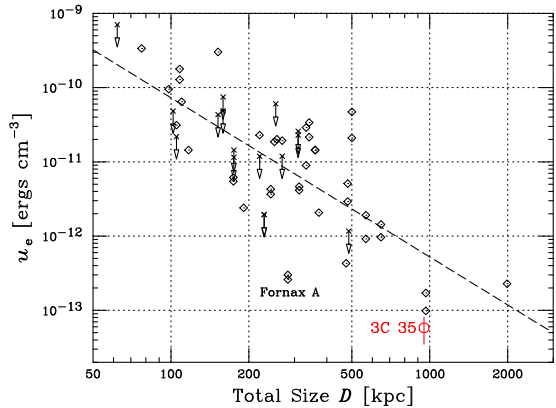}
\end{center}
\caption{$u_{\rm e}$ plotted against the total size $D$ of radio galaxies.
The circle corresponds to the data of 3C 35,
with only the statistical error in $S_{\rm 1keV}$ plotted.
Diamonds shows lobes from which the IC X-rays were securely detected,
while crosses with a downward arrow 
indicate those with only the upper limit on the IC flux
\citep[][and reference therein]{3C326}.
The dashed line represents the best-fit relation 
of $u_{\rm e} \propto D^{-2.1}$,
for the lobes displayed with the diamonds.
}
\label{fig:size2energetics}
\end{figure}

In Figure \ref{fig:size2energetics}, \ue~in the lobes is plotted 
against the total linear size $D$ of the radio galaxies. 
As is pointed out by \citet{3C326},
the figure suggests a correlation of $u_{\rm e} \propto D^{-2.1 \pm 0.3 }$ 
in the range of $ D = 50$ kpc -- $2$ Mpc 
(the dashed line in Figure \ref{fig:size2energetics}).
In order to derive this relation, 
the upper limits 
(the crosses with an arrow in Figure \ref{fig:size2energetics})
were neglected. 
This relation is indicated by the dashed line 
in Figure \ref{fig:size2energetics}.
It was proposed that the radio galaxies follow this trend
while their jets are actively injecting sufficient energy into their lobes,
based on the following argument \citep{3C326}.
The size of the radio sources is reported to scale roughly as $D \propto \tau$ 
over a wide range of $D = 10$ pc -- $10^3$ kpc \citep[][]{size2age},
where $\tau$ is the age of the radio galaxy and its jet.
Assuming that the energy input rate is constant in time, 
the total energy input to the lobes, $E$, 
is considered to be in proportion to $\tau$,
before radiative cooling becomes dominant. 
These simple assumptions give 
the relation of $u_{\rm e} \sim E/V \sim D^{-2}$.
In contrast, 
after the jets become inactive and cease the energy transport, 
\ue~in the lobes should decrease due to the adiabatic expansion of the lobes, 
and to the synchrotron and IC radiation energy losses.
Correspondingly, the data points for such sources should deviate downward 
from the \ue-$D$ correlation in Figure \ref{fig:size2energetics}.
Actually, in the lobes of Fornax A 
which is reported to host a dormant nucleus \citep{ForA_nucleus,ForA_chandra},
the values of \ue, precisely determined 
through various X-ray observations \citep[e.g.,][]{ForA,ForA_Suzaku}, 
are found to be lower than the trend of the other radio galaxies
by one order of magnitude (see Figure \ref{fig:size2energetics}). 

%==========% 
% Figure 7 %
%==========%

Radio images \citep[e.g.,][]{3C35_RadioImage,3C35_RadioIndex} 
have revealed the hot spots in the individual lobes of \src,
suggesting that these lobes are still energized by the jets. 
In spite of this, 
the data point for the lobes of \src~deviates below by a factor of $\sim 10$
from the \ue-$D$ regression curve in Figure \ref{fig:size2energetics},
similar to Fornax A. 
A possible scenario to settle this apparent discrepancy is that 
the current jet power to the lobes of \src~has become significantly lower 
than that in the past, as is expected from 
the smaller and hence younger radio sources with a size of $\sim 100$ kpc.
This idea seems to be supported by the observational fact that 
no jet feature was detected from the nucleus of \src~\citep{RG_summary4},
even though about $80$\% of LERGs exhibited definite/possible jets.
This is also compatible with the inactive X-ray nucleus of \src~as a LERG
with a 0.5 -- 10 keV luminosity of $\ll 10^{42}$ \lum,
as is discussed in \S \ref{sec:chandra_src}.
Another interpretation is given 
by a model for dynamical evolution of a radio source 
\citep[e.g.,][]{Evolution_on_PDdiagram},
proposed to interpret the behavior of radio sources 
on the radio power-size diagram (so-called the $P$-$D$ diagram),
on which a deficit of luminous giant radio sources is observed. 
The model predicts that 
the power of a radio source decreases suddenly on the $P$-$D$ diagram 
due to the dominance of the IC loss 
without assuming the decline of the jet activity,
when the source size reaches $D\sim 1$ Mpc. 
This prediction qualitatively appears to consistent with the low value of \um,
observed in the lobes of \src. 
The current X-ray knowledge on the giant radio galaxies
is insufficient to distinguish these two possibilities. 
A systematic X-ray study on giant radio galaxies is necessary.

%===========% 
% Figure A1 %
%===========%
\begin{figure}[b]
\begin{center}
\includegraphics[angle=-90,width=8cm]{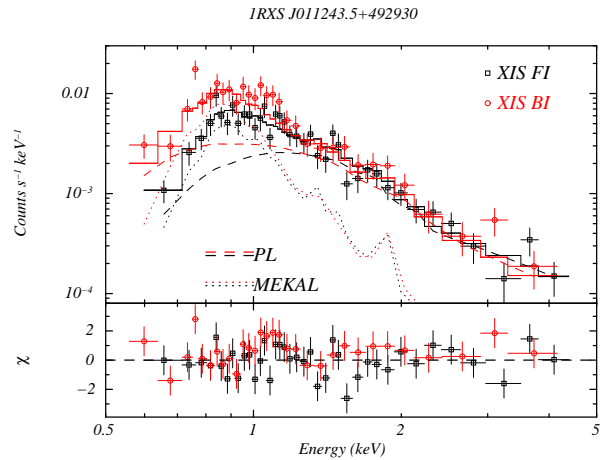}
\end{center}
\caption{The XIS spectrum of \rxs. 
The best-fit PL+MEKAL model is shown with the histograms.}
\label{fig:RXS}
\end{figure}

%----------%
% Table A1 %
%----------%
\begin{table}[t]
\caption{Best-fit spectral parameters for \rxs}
\label{table:1RXS}
\begin{tabular}{lll}
\hline  \hline %===================================================
Parameters                      &  PL                 & PL+MEKAL    \\
\hline %------------------------------------------------
$N_{\rm H}$ ($10^{21}$ \colden)  & $2.1_{-0.8}^{+0.9}$    & $1.5_{-0.7}^{+0.8}$\\
$\Gamma$                        & $3.6_{-0.4}^{+0.6}$   & $2.6_{-0.3}^{+0.4}$ \\
$kT$ (keV)                      & --                   & $0.62 \pm 0.06 $ \\
$F_{\rm X}$ ($10^{-13}$ \flux)\tablenotemark{a}
                                & $2.6 \pm 0.1 $       & $2.7_{-0.3}^{+0.2}$\\
$\chi^2/{\rm d.o.f}$            & $131.6/67$           & $74.5/65$    \\ 
\hline %------------------------------------------------
\end{tabular}
\tablenotetext{a}{The total observed X-ray flux in 0.5 -- 5 keV.}
\end{table}

\acknowledgments
Thanks to the supportive advices from the anonymous reviewer,
the paper was significantly improved.  
This research was made possible, 
owing to the successful operation and calibration of \suzaku.
This research made use of the \chandra~data 
and associated softwares provided by the Chandra X-ray Center.
The support is acknowledged from the Ministry of Education, Culture, Sports, 
Science and Technology (MEXT) of Japan 
through the Grant-in-Aid for the Global COE Program,
"The Next Generation of Physics, Spun from Universality and Emergence".
N. I. is supported by the Grant-in-Aid for Young Scientists (B) 22740120 
from the MEXT.

\appendix
\section{\suzaku~spectrum of \rxs}
The XIS spectrum of \rxs~\citep{1RXS},
the brightest source within the XIS FoV, is briefly analyzed,
although detailed discussions are beyond the scope of the present paper. 
The XIS events were extracted from a circle with a radius of $1'.5$ 
centered on the source (the white cross in Figure \ref{fig:image}),
while the background (NXB+XRB) events were simply taken 
from a neighboring source free circular region with a $2'$ radius. 
The background-subtracted XIS spectrum of \rxs~in the 0.5 -- 5 keV range
is shown in Figure \ref{fig:RXS}.

The spectrum appears to be rather soft. 
A simple PL model subjected to a free absorption failed 
to describe the observed spectrum ($\chi^2/{\rm d.o.f} = 131.6/67$). 
By adding a MEKAL component with the solar abundance, 
the fit became acceptable
with the parameters listed in Table \ref{table:1RXS}.
The spectrum gave the MEKAL temperature and PL photon index 
of $kT = 0.62 \pm 0.06 $ (keV) and  $\Gamma = 2.6_{-0.3}^{+0.4}$, respectively.
The absorption-inclusive X-ray flux was measured 
as $2.7_{-0.3}^{+0.2} \times 10^{-13}$ \flux~in 0.5 -- 5 keV,
although the intrinsic source luminosity can not be evaluated, 
due to the unknown redshift of the source.

%%%%%%%%%%%%%%%
% References %
%%%%%%%%%%%%%%%

%%%%%%%%%%%%%%%%% Tables %%%%%%%%%%%%%%%%%

%%%%%%%%%%%%%%%%% Figures %%%%%%%%%%%%%%%%%

\end{document}